\documentclass[twocolumn,pre,floatfix]{revtex4}
\usepackage{graphicx}

\newsavebox{\mystdbox} 
\sbox{\mystdbox}{$\scriptstyle-$}
\newlength{\mystdlen}
\newcommand{\mystd}{%
  {\makebox[0pt][l]{\makebox[\mystdlen][c]{$\scriptstyle\circ$}}%
  \usebox{\mystdbox}}%
}

\newcommand{\rhoistd}{\rho_i^\mystd}
\newcommand{\kB}{k_{\mathrm{B}}}
\newcommand{\kT}{\kB T}
\newcommand{\lB}{l_{\mathrm{B}}}
\newcommand{\dthree}{d^3}
\newcommand{\myvector}[1]{\mathbf{#1}}
\newcommand{\rvec}{\myvector{r}}
\newcommand{\qvec}{\myvector{q}}
\newcommand{\rhoz}{\rho_z}
\newcommand{\fself}{f_m}
\newcommand{\vf}{\phi}
\newcommand{\pot}{\psi}
\newcommand{\omegaback}{\omega(\pm\infty)}

\newcommand{\Fex}{F_{\mathrm{ex}}}
\newcommand{\loge}{\ln}
\newcommand{\nm}{\mathrm{nm}}
\newcommand{\muM}{\mu\mathrm{M}}

\newcommand{\PB}{Poisson-Boltzmann}
\newcommand{\DebH}{Debye-H\"uckel}
\newcommand{\SL}{Stillinger-Lovett}
\newcommand{\Rescic}{Re\v{s}\v{c}i\v{c}}
\newcommand{\Lif}{Lifshitz}
\newcommand{\FW}{Fisher-Widom}
\newcommand{\KFW}{Kirkwood-\FW}

\newcommand{\eqref}[1]{(\ref{#1})}

\begin{document}

\title{Density functional theory for a macroion suspension}

\author{Patrick B. Warren}
\affiliation{Unilever R\&D Port Sunlight, Bebington, Wirral, CH63 3JW, UK.}

\date{June 9, 2005}

\begin{abstract}
A density functional theory for a macroion suspension is examined,
where the excess free energy corresponds to the macroion self energy
arising from the polarisation of the supporting electrolyte solution.
This is treated within a linearised or Debye-H\"uckel approximation.
The model predicts liquid-liquid phase separation at low ionic
strength.  The interface structure and surface tension between
coexisting phases is calculated using a variational approximation.
Results are also obtained for structure factors, which are shown to
obey the Stillinger-Lovett moment conditions.  As one approaches the
critical points, the structure factors may diverge at a non-zero
wavevector, indicating that the critical points could be replaced by
charge-density-wave phases.
\end{abstract}

\maketitle

\section{Introduction}
\label{sec:intro}
The phase behaviour of charged colloidal suspensions at low ionic
strength has attracted much experimental and theoretical interest.
Observations of void structures and other phenomena \cite{phenomena,
GrAnt} motivated a number of theoretical studies which attributed the
anomalous behaviour to phase separation between colloid-rich and
colloid-poor phases \cite{theory, vRH2, Warren1, Chan1}.  Several
reviews are available \cite{reviews, Bell-rev}.  The original
theoretical explanations have come under strong attack for using a
\DebH\ linearisation approximation which is, at best, at the margin of
its validity.  Various attempts to patch this up have left the
situation unclear.  Cell model calculations using \PB\ theory indicate
the original predictions are an artefact of the linearisation
approximation \cite{GvRKDesG, TamS}.  The \DebH\ approximation can be
improved by taking into account counterion condensation, in which case
the phase transition may or may not be recovered depending on the
approximation scheme used \cite{dhfixes}.  Other approaches such as
extended \DebH\ theory \cite{CLPet}, symmetrised \PB\ theory
\cite{BOuth1, BOuth2}, `boot-strap' \PB\ theory \cite{PetCL}, and a
systematic expansion into two- and three-body interactions
\cite{threebody, HDRoij}, all indicate that a phase transition can
occur, as do several integral equation studies \cite{inteqs, GonzTo}.
The experimental situation is also uncertain since a plausible
alternative explanation has been suggested \cite{Bell-rev}, in which
the voids correspond to regions occupied by dilute, highly extended
(and therefore effectively invisble) polyelectrolyte chains which have
been shed by the latex colloids.

Simulation methods struggle to approach these problems because of the
disparity in size between the macroions and the small ions, and the
need to handle the electrostatic interactions.  Nevertheless,
convincing evidence has been found for liquid-liquid phase separation
in a macroion system at lower dimensionless temperatures
\cite{simulations, ResLin}.  Experimentally this corresponds to a
solvent with a lower dielectric constant than water (but one in which
the ions still disperse).  Charged colloids in such solvents exhibit
many interesting phenomena \cite{nonaqueous}.

Thus, whilst the weight of evidence perhaps suggests that aqueous
charge-stabilised colloidal suspensions may not show genuine
liquid-liquid phase coexistence, it is absolutely clear that there
will be phase coexistence between condensed and dilute colloidal
phases at small enough dimensionless temperatures.  In this sense, the
problem resembles the much-studied restricted primitive model (RPM),
whose phase behaviour is now well established \cite{Fisher, Stell,
LFishP, KFishL}.

In situations where genuine phase coexistence obtains, one can go on
to ask questions about the surface tension and electrical structure of
the interface between the coexisting phases.  Answers to these
questions may prompt new avenues for experimental investigation of
real systems.  Previously, Knott and Ford compute the surface tension
using square-gradient theory, but discard the possible electrical
structure at the interface \cite{KnottF}.  The present work approaches
this problem within the context of a density functional theory,
motivated by the earlier study in Ref.~\cite{Warren1} (see also
Appendix A).  It places the phenomenological remarks made
in this earlier work on a sounder footing.  The analysis in
Ref.~\cite{Warren1} suggests that the macroion self energy is the
dominant contribution to the excess free energy, similar to an early
insight by Langmuir \cite{Langmuir}.  In the present work therefore,
the rather gross simplification has been adopted in which the macroion
self energy is the \emph{only} contribution to the excess free energy.
Moreover this self energy is computed in a simple closed form using
\DebH\ theory, and is thus also based on the much-criticised
linearisation approximation.  Nevertheless I argue that it is
instructive to proceed, because of the rich phenomenology that is
revealed.

The model predicts phase separation at low dimensionless temperatures
and low ionic strengths, and in quantitative terms stands reasonable
comparison with some of the other approaches.  The physics of the
phase separation lies in the dependence of the macroion self energy on
the local ionic strength: macroions drift towards regions of high
ionic strength, which by charge neutrality are regions where other
macroions have also congregated.  Within the linearisation
approximation, the effect grows without bound as the macroion charge
is increased, and thus the mechanism can drive phase separation at
sufficiently large macroion charges.  In reality, non-linear effects
(counterion condensation) limit the effective macroion charge
\cite{saturation}, and therefore this
mechanism is probably insufficient in itself to drive phase separation
in real systems.  Undoubtably though it is still a contributing
factor, operating in conjunction with other effects such as correlated
fluctuations in the counterion clouds around macroions and the sharing
of counterions between macroions \cite{Schmitz3, GrAnt, simulations}.

The model is constructed in the form of a density functional theory.
Thus, as well as making predictions for phase separation, it can be
used to solve for the density profiles and the surface tension between
coexisting phases.  The results obtained here are in accord with
typical expectations for soft condensed matter systems
\cite{surftens}, and were summarised in an earlier publication
\cite{Warren2}.  In addition, I also discuss the predictions that the
theory makes for the structure factors.  These are found to obey the
Stillinger-Lovett moment conditions \cite{StillLov, Martin}, although
it turns out this is not a stringent test of the theory.
Intriguingly, I find that the structure factors may diverge at a
non-zero wavevector as one approaches the critical points.  This
suggests the possibility that the critical points in these systems may
be replaced by charge-density-wave phases \cite{Warren3}.  This
phenomenological possibility in charged systems was first suggested by
Nabutovskii and coworkers \cite{NabuNP1, HStell, Fisher}.

\section{Specification of the model}
\label{sec:model}
The underlying model of the macroion suspension used here is a
primitive model commonly deployed for this kind of problem.  The
`primitive' aspect is that the solvent is treated as a structureless
dielectric continuum in which the macroions and small salt ions are
embedded.  The macroions are treated as spheres of (positive) charge
$Z$, diameter $\sigma$, and number density $\rho_m$ (volume fraction
$\vf=\pi\sigma^3\rho_m/6$).  The salt ions are univalent counterions
and coions at number densities $\rho_-$ and $\rho_+$ respectively.  I
suppose there is only one species of counterion.  The size of the salt
ions is assumed to be small enough to be irrelevant.  The dielectric
continuum is characterised by a Bjerrum length $\lB$ so that the
electrostatic interaction energy between a pair of univalent charges
separated by a distance $r$ is $\lB/r$, in units of $\kT$ where $\kB$
is Boltzmann's constant and $T$ is the temperature.  For water at room
temperature, $\lB\approx0.72\,\nm$.  The model is completely
parametrised by the dimensionless ratio $\sigma/\lB$ and the charge
$Z$.  It is often convenient to pretend that the dielectric
permittivity of the background is independent of temperature, in which
case $\lB\sim1/T$.  This means that $\sigma/\lB$ can be regarded as
a dimensionless temperature.

The density functional theory is specified by giving the free energy
$F$ as a functional of the spatially varying number densities
$\rho_m(\rvec)$ and $\rho_\pm(\rvec)$ \cite{EvansDFT}.  The
functional is decomposed into ideal, mean-field, and correlation
contributions:
\begin{equation}
\begin{array}{l}
\displaystyle\frac{F}{\kT}=\int\!\dthree\rvec\sum_{i=m,\pm}
\rho_i(\rvec)\,\loge\frac{\rho_i(\rvec)}{e\rhoistd}\\[18pt]
\displaystyle\hspace{5em}{}+\frac{\lB}{2}\int\!\dthree\rvec\, 
\dthree\rvec'\,
\frac{\rhoz(\rvec)\rhoz(\rvec')}{|\rvec-\rvec'|},\\[18pt]
\displaystyle\hspace{10em}{}+\frac{1}{\kT}
\int\!\dthree\rvec\,\rho_m(\rvec)\,\fself(\rvec).
\end{array}
\label{eq:dft}
\end{equation}
The first term is the ideal term: $e$ is the base of natural
logarithms and the $\rhoistd$ are unimportant base units of
concentration related to the definition of the standard state
\cite{EBSmith}.  The second term is a mean-field electrostatics term:
$\rhoz(\rvec) = \sum_i z_i\rho_i(\rvec)$ is the local charge density
with $z_i=\{Z,1,-1\}$ as $i=\{m,+,-\}$, and a factor $1/2$ allows for
double counting.  The third term (correlation term) represents the
excess free energy.  As discussed above, only the macroion self
energy $\fself$ is included in this term.  This is computed using \DebH\
theory \cite{vRH1, vRH2, BCM, note1},
\begin{equation}
\fself(\rvec)=\frac{2Z^2\lB\kT}{\sigma(\sigma\kappa(\rvec)+2)},
\label{eq:fm}
\end{equation}
where $\kappa(\rvec)$ is a local inverse Debye screening length.  This
is defined in terms of an \emph{average} local ionic strength,
$\overline\rho_I(\rvec)$, through
\begin{equation}
\begin{array}{l}
\displaystyle[\kappa(\rvec)]^2=8\pi\lB\overline\rho_I(\rvec),\\[6pt]
\displaystyle\overline\rho_I(\rvec)={\textstyle\int}\dthree\rvec'\,
w(|\rvec-\rvec'|)\,\rho_I(\rvec')\\[6pt]
\displaystyle\rho_I(\rvec') = [\rho_+(\rvec') + \rho_-(\rvec')]/2.
\end{array}
\label{eq:rhoi}
\end{equation}
The ionic strength includes the counterions and salt ions, but not the
macroions.  In principle, allowance should be made for the macroion
excluded volume, but this effect is of secondary importance and for
simplicity has been omitted.

The smoothing kernel in the second of Eqs.~\eqref{eq:rhoi} is
normalised so that $\int\dthree\rvec\,w(r)=1$.  Here I use
\begin{equation}
w(r)=(\pi\alpha\sigma^2)^{-3/2}\,\exp[-r^2/(\alpha\sigma^2)].
\label{eq:wr}
\end{equation}
This is an arbitrarily chosen function \cite{note1b}, of range
$\alpha^{1/2}\sigma$.  The argument below suggests that the parameter
$\alpha$ should be of order unity and for the most part I will set
$\alpha=1$ in the calculations.  Eqs.~\eqref{eq:dft}--\eqref{eq:wr}
completely specify the density functional theory, and everything
discussed below can be derived from them.

The decomposition into ideal, mean field, and correlation
contributions is a standard approach \cite{EvSl, Sluckin,
decomposition}.  The approximation made for the correlation term
deserves more discussion though.  The only piece of physics that has
been incorporated is the macroion self energy.  This has a non-trivial
dependence on the local ionic strength since each macroion polarises
the surrounding electrolyte and becomes surrounded by a `double
layer'.  This dependence causes macroions to drift towards regions of
high ionic strength, as discussed already.

The physical reason for introducing a smoothing kernel is that one can
derive the self energy by integrating out the small ion degrees of
freedom, with the main contribution coming from variations on length
scales corresponding to the structure in the double layer \cite{vRH2}.
Thus only variations in ionic strength on length scales $\agt\sigma$
should be included in the model. The smoothing kernel is a device for
achieving this.  This argument also motivates the choice for $\alpha$
in Eq.~\eqref{eq:wr}.

In section \ref{sec:sf} below, it is found that the theory is not well
behaved if one uses a `point model' where the dependence is on the
ionic strength at, say, the centre of the macroion (the first of
Eqs.~\eqref{eq:rhoi} with $\overline\rho_I$ replaced by $\rho_I$).
This provides a second technical reason to make the self energy depend
on a smeared ionic strength.

The potential energy of a small ion at the surface of the macroion, in
units of $\kT$, is $\pm Z\lB/\sigma$.  Eq.~\eqref{eq:fm} uses the
\DebH\ expression for the self energy, which assumes
$Z\lB/\sigma\ll1$.  The expression becomes increasingly inaccurate for
$Z\lB/\sigma\agt1$, and its use has been the subject of strong
criticism as discussed above.  Since the interesting effects are found
only at larger values of $Z\lB/\sigma$, one should interpret the
quantitative results with caution.

\begin{figure}
\begin{center}
\includegraphics{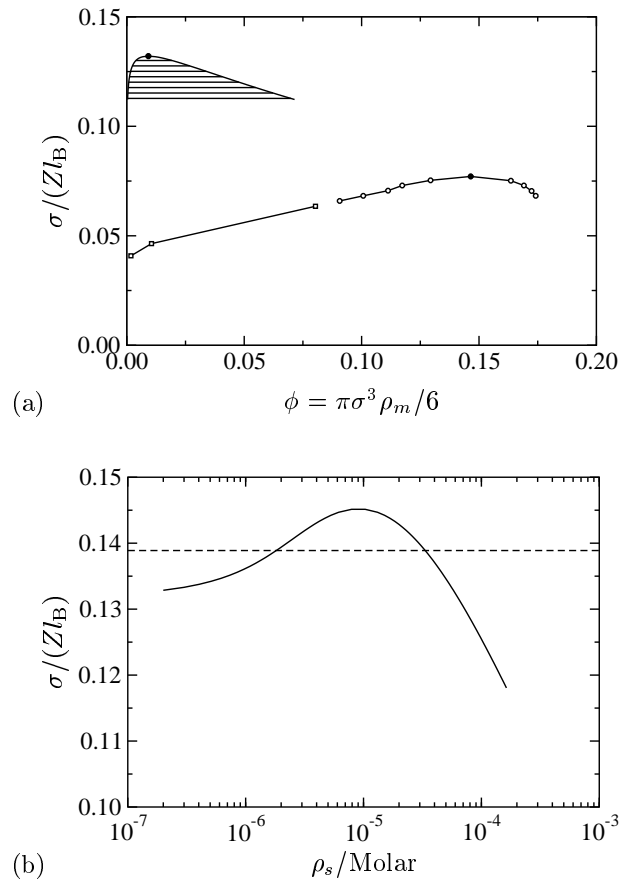}
\end{center}
\caption[?]{(a) Universal phase behaviour in the absence of salt,
predicted by Eq.~\eqref{eq:nosalt} (upper curve).  The simulation
results of \Rescic\ and Linse \cite{ResLin} are also shown (lower
curve with marked points). (b) Behaviour of the critical point at
$Z=10^3$ as salt is added.  The dashed line corresponds to the
parameters used in Fig.~\ref{fig:bino}.\label{fig:simhyp}}
\end{figure}

\section{Bulk phase behaviour}
\label{sec:bulk}
In this section, I shall consider the bulk phase behaviour predicted
by the free energy of Eqs.~\eqref{eq:dft}--\eqref{eq:wr}.  This is a
homogeneous situation in which the density variables lose their
spatial dependence.  In this limit, one can prove that the mean field
term should be replaced by a condition of bulk charge neutrality,
$\rhoz=\sum_iz_i\rho_i=0$ \cite{Warren0, TamS}.

The required charge neutrality condition can be imposed in two ways.
The first route is to add a term $\pot\kT\sum_iz_i\rho_i$ to the free
energy, where $\pot\kT$ is a Lagrange multiplier.  This approach has
the advantage of making a close connection to the density functional
theory.  Taking this approach, the free energy becomes
\begin{equation}
\frac{F}{V\kT}=\sum_i\rho_i\Bigl(
\loge\frac{\rho_i}{e\rhoistd}+z_i\pot\Bigl)
+\frac{2Z^2\lB\rho_m}{\sigma(\sigma\kappa+2)}
\label{eq:bfe}
\end{equation}
where $V$ is the system volume and $\kappa^2 = 4\pi\lB(\rho_+ +
\rho_-)$.  The distinction between the smoothed and unsmoothed ionic
strength disappears in the homogeneous limit.  In this approach the
$\rho_i$ are treated as three independent density variables.  At the
end of any calculations, $\pot$ is adjusted to get $\sum_iz_i\rho_i =
0$.  The value of $\pot$ depends on the state point under
consideration.

The second way to enforce charge neutrality is to eliminate one of the
density variables.  Since this is numerically quite convenient, it is
the approach that shall be adopted in the rest of this section.  At
this point one can recognise that the coions come from added salt and
write $\rho_-=Z\rho_m+\rho_s$ and $\rho_+=\rho_s$, where $\rho_s$ is
the added salt concentration.  The free energy is given by
Eq.~\eqref{eq:bfe} but with $\pot=0$, and $\rho_{\pm}$ substituted by
the above expressions.  There are now only two independent density
variables and the phase behaviour can be represented in the $(\rho_m,
\rho_s)$ plane.
 
I now discuss the phase behaviour predicted by this free energy.
Firstly, in the absence of salt some additional simplifications can be
made.  In the limit $\rho_s\to0$, the free energy can be written in
a dimensionless form as
\begin{equation}
\frac{\pi\sigma^3 F}{6ZV\kT}=
\vf\loge{\vf}+\frac{2\vf Z \lB/\sigma}
{(24\vf Z\lB/\sigma)^{1/2}+2}\label{eq:nosalt}
\end{equation}
where $\vf$ is the macroion volume fraction.  To get to this point, I
have assumed that $Z\gg1$ and hidden some constants and terms strictly
proportional to $\rho_m$ since they do not affect the phase behaviour.

Eq.~\eqref{eq:nosalt} predicts the dependence on $\sigma/\lB$ and $Z$
is through the single combination $Z\lB/\sigma$ (there is no reason to
suppose that this should be the case in a more accurate theory).  This
is the same parameter that quantifies the accuracy of the \DebH\
linearisation approximation.  The inverse of this, $\sigma/(Z\lB)$, is
proportional to the dimensionless temperature discussed above.
Fig.~\ref{fig:simhyp}(a) shows the universal phase behaviour predicted
by Eq.~\eqref{eq:nosalt} as a function of the macroion volume fraction
and $\sigma/(Z\lB)$.  At small enough values of $\sigma/(Z\lB)$, a two
phase region is encountered in the phase diagram.  The two phase
region corresponds to phase coexistence between macroion rich and
macroion poor phases.  The identities of these phases merge at a
critical point located at $\vf \approx 9.18\times10^{-3}$ and
$\sigma/(Z\lB) \approx 0.132$.

One can compare this with the simulation results of \Rescic\ and Linse
for $Z=10$ macroions \cite{ResLin}.  They also find a two phase region
on lowering temperature, with a critical point located at $\vf \approx
0.17$ and $\sigma/(Z\lB) \approx 0.077$.  Whilst the phenomenology is
the same, the numerical values are somewhat different from the
prediction of Eq.~\eqref{eq:nosalt}.  Not unexpectedly, the present
model is too crude to obtain quantitatively reliable
results.  An analogy can be made with the application of \DebH\ theory
to the restricted primitive model (RPM) \cite{FishLev, Fisher,
LFishP}. In this case too, \DebH\ theory correctly suggests a region
of phase separation at low temperatures but errs in terms of
quantitative predictions.  Interestingly, in terms of accuracy of
prediction, the present theory is not much worse than symmetrised \PB\
theory or the mean spherical approximation \cite{GonzTo, BOuth1,
note2b}.

\begin{figure}
\begin{center}
\includegraphics{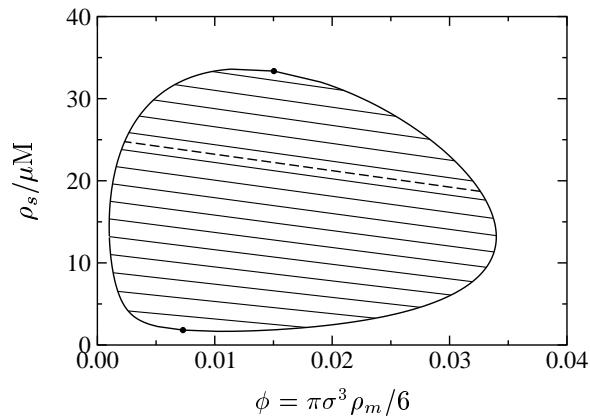}
\end{center}
\caption[?]{Phase behaviour at $Z=10^3$, $\sigma=100\,\nm$ and
$\lB=0.72\,\nm$, corresponding to the dashed line in
Fig.~\ref{fig:simhyp}.  The miscibility gap is bounded above and below
by critical points.  The dashed tie line is the one for which the
interfacial properties are reported in
Figs.~\ref{fig:ionprof}--\ref{fig:omprof}.\label{fig:bino}}
\end{figure}

I now turn the effect of added salt, and analyse the predictions of
the full free energy in Eq.~\eqref{eq:bfe}.  In general, as salt is
added, the critical point in Fig.~\ref{fig:simhyp}(a) first moves to
higher dimensionless temperatures, passes through a maximum, and then
starts to move to lower dimensionless temperatures again.  This
non-monotonic behaviour is shown in Fig.~\ref{fig:simhyp}(b) for
$Z=10^3$.  A similar effect of added salt is seen in a number of other
approaches \cite{vRH1, vRH2, Warren1, DBLev, PetC, DWHans}.  In the
presence of added salt, it is no longer true that the dependence on
$Z$ and $\sigma/\lB$ can be combined into a single parameter, however
for comparison with the phase behaviour in the absence of salt,
Fig.~\ref{fig:simhyp}(b) shows the behavior as a function of
$\sigma/(Z\lB)$ at this fixed value of $Z$.

The re-entrant behaviour means that for parameters such as those
corresponding to the dashed line in Fig.~\ref{fig:simhyp}(b), there
are \emph{two} critical points in the $(\rho_m, \rho_s)$ plane, and one
encounters a re-entrant single phase region at low added salt.  The
dashed line in Fig.~\ref{fig:simhyp}(b) is for $Z=10^3$,
$\sigma=100\,\nm$ and $\lB=0.72\,\nm$, and the corresponding phase
behaviour in the $(\rho_m, \rho_s)$ plane is shown in Fig.~\ref{fig:bino}.
It is seen that the two phase region appears as a miscibility gap in
this representation.

As $\sigma/\lB$ is increased or $Z$ is decreased, the two critical
points move towards each other and finally disappear at a double
critical point, or hypercritical point \cite{WalkV}.  For example, for
$Z=10^3$ the double critical point corresponds to the maximum of the
solid line in Fig.~\ref{fig:simhyp}(b), where $\sigma/(Z\lB) \approx
0.145$, $\vf \approx 1.04\times10^{-2}$ and $\rho_s \approx
8.98\,\muM$.

The bulk phase behaviour predicted by Eq.~\eqref{eq:bfe} thus closely
resembles that predicted by various other approaches, including the
theory discussed in Ref.~\cite{Warren1}.  Many approaches, including
the present one, do not consider the formation of ordered phases
(colloidal crystals).  These can arise from the strong
macroion-macroion interactions.  The possibility of ordered phases has
been considered by van Roij and coworkers \cite{vRH1, vRH2, HDRoij}
though.  They find that ordered phases can appear in the vicinity of
the miscibility gap in which case a richer phase behaviour can result.

\begin{figure}
\begin{center}
\includegraphics{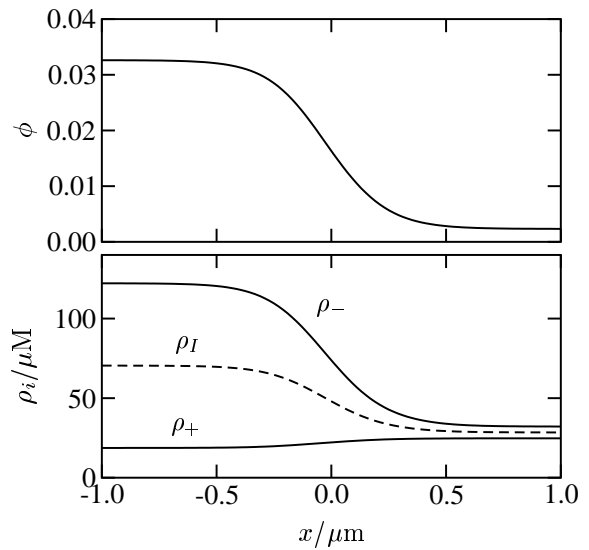}
\end{center}
\caption[?]{Macroion volume fraction (top) and small ion
concentrations (bottom) through the interface corresponding to the
dashed tie line in Fig.~\ref{fig:bino}.\label{fig:ionprof}}
\end{figure}

\begin{figure}
\begin{center}
\includegraphics{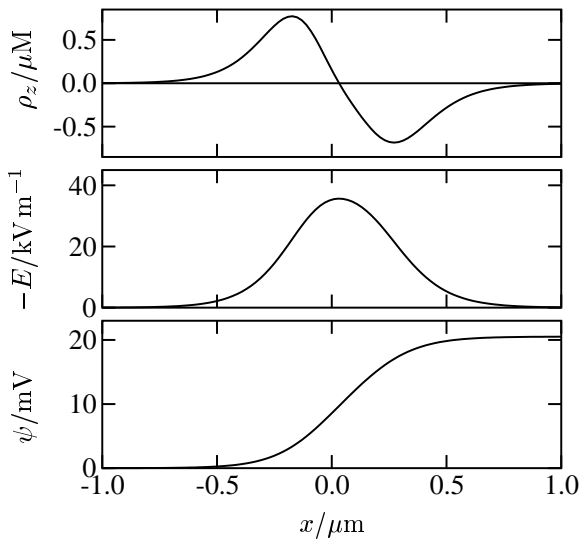}
\end{center}
\caption[?]{Charge density (top), electric field (middle), and
electrostatic potential (bottom) corresponding to the ion density
profiles shown in Fig.~\ref{fig:ionprof}.\label{fig:fieldprof}}
\end{figure}

\section{Interfacial properties}
\label{sec:profile}
A major use of the density functional theory in the present context is
to calculate the macroion and small ion density profiles through the
interface between two coexisting phases, and to compute the surface
tension.  In order to set the problem up, it is convenient to
introduce the grand potential \cite{EvansDFT}
\begin{equation}
\Omega=F-\int\!\dthree\rvec\sum_{i=m,\pm}\mu_i\rho_i(\rvec)
\label{eq:omega}
\end{equation}
where $\mu_i$ are the chemical potentials of the three species, and
$F$ is defined in Eqs.~\eqref{eq:dft}--\eqref{eq:wr}.  At this point
it is also convenient to rewrite the mean field term in
Eq.~\eqref{eq:dft}.  Define a dimensionless electrostatic potential
\begin{equation}
\pot(\rvec)=\lB \int\!\dthree\rvec'\,
\frac{\rhoz(\rvec')}{|\rvec-\rvec'|}\label{eq:potdef}
\end{equation}
so that the mean field term in Eq.~\eqref{eq:dft} can be written
\begin{equation}
\frac{\lB}{2}\int\!\dthree\rvec\, \dthree\rvec'\,
\frac{\rhoz(\rvec)\rhoz(\rvec')}{|\rvec-\rvec'|}
= \frac{1}{2}\int\!\dthree\rvec\,\pot(\rvec)\,\rhoz(\rvec).\label{eq:pd2}
\end{equation}
By direct substitution, one verifies that the potential defined by
Eq.~\eqref{eq:potdef} solves the Poisson equation
\begin{equation}
\nabla^2\pot + 4\pi\lB\rhoz = 0.\label{eq:poisson}
\end{equation}
Using this and Green's first identity \cite{Hinchey}, the mean field
term now becomes
\begin{equation}
\frac{1}{2}\int\!\dthree\rvec\,\pot(\rvec)\,\rhoz(\rvec)
=\frac{1}{8\pi\lB}\int\!\dthree\rvec\,|\nabla\pot|^2.
\end{equation}
This is recognised as the electric field energy since $\nabla\pot$ is
essentially the electric field strength.  One can now define a grand
potential density $\omega(\rvec)$ such that $\Omega =
\int\!\dthree\rvec\,\omega(\rvec)$ and
\begin{equation}
\omega=\sum_{i}
\rho_i\Bigl(\kT\loge\frac{\rho_i}{e\rhoistd}-\mu_i\Bigr)
+\frac{\kT}{8\pi\lB}|\nabla\pot|^2
+\fself\rho_m\label{eq:omdens}
\end{equation}
where the explicit dependence on the spatial co-ordinate has been
suppressed.  For a homogeneous system, $\omega=-p$ where $p$ is the
pressure.

\begin{figure}
\begin{center}
\includegraphics{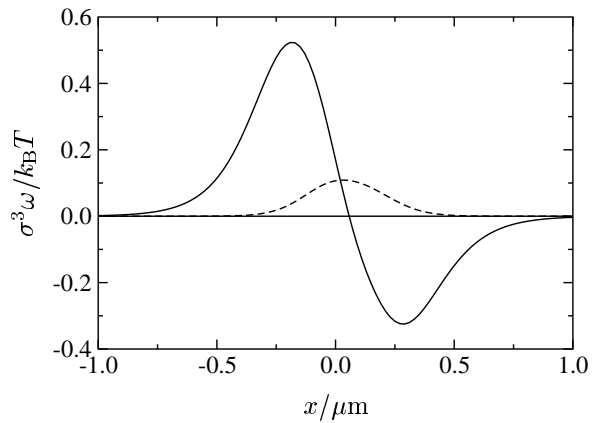}
\end{center}
\caption[?]{Excess grand potential density (solid line) and
electrostatic component thereof (dashed line) corresponding to the ion
density profiles shown in Fig.~\ref{fig:ionprof}.\label{fig:omprof}}
\end{figure}

Setting $\delta\Omega/\delta\rho_i(\rvec)=0$ and
using Eq.~\eqref{eq:potdef} gives
\begin{equation}
\begin{array}{l}
\displaystyle\frac{\mu_i}{\kT}=
\loge\frac{\rho_i(\rvec)}{\rhoistd}+z_i\pot(\rvec)\\[18pt]
\displaystyle\hspace{5em}{}+\frac{\delta}{\delta\rho_i(\rvec)}\Bigl(
\frac{\textstyle\int\!\dthree\rvec'\,\rho_m(\rvec')\,
\fself(\rvec')}{\kT}\Bigr).
\end{array}
\label{eq:horrid}
\end{equation}
In principle, these non-linear integral equations can be solved to
find the ion density profiles.  Here a variational approximation has
been adopted in which $\Omega$ is minimised with respect to parameters
in trial functions which specify the ion density profiles.  More
details of the numerical approach are given in Appendix B.

I now suppose that all the variation occurs in one direction $x$
normal to the interface.  At large distances from the interface,
$x\to\pm\infty$, the number densities approach those corresponding to
the coexisting bulk phases.  The grand potential density approaches a
constant value $\omegaback$ equal to (minus) the pressure, and
therefore the same in coexisting phases.  The surface tension $\gamma$
can therefore be identified as the excess grand potential per unit
area
\begin{equation}
\textstyle\gamma=\int_{-\infty}^\infty dx\,[\omega-\omegaback].
\label{eq:gamma}
\end{equation}

The chemical potentials derived from Eq.~\eqref{eq:bfe} are
\begin{equation}
\frac{\mu_i}{\kT}=\loge\frac{\rho_i}{\rhoistd}+z_i\pot
+\frac{\partial}{\partial\rho_i}\Bigl(
\frac{2Z^2\lB\rho_m}{\sigma(\sigma\kappa+2)}\Bigr).
\label{eq:mu}
\end{equation}
Comparison with Eq.~\eqref{eq:horrid} shows that $\pot$ in this
expression is simply the limiting value of $\pot(\rvec)$ in the case
of a homogeneous system \cite{note2}.  For the interface problem, one
has two limiting values, $\pot(\pm\infty)$.  The difference
$\Delta\pot=\pot(\infty)-\pot(-\infty)$ arises because of the
electrical structure at the interface.  It is a liquid-liquid junction
potential analogous to the Donnan potential that appears across a
semi-permeable membrane \cite{Donnan}.  Since $\pot$ in
Eq.~\eqref{eq:mu} is determined by the bulk densities, the difference
$\Delta\pot$ can be calculated without having to solve for the interface
structure.  In fact, because of the symmetric way that $\rho_{\pm}$
enters into the excess free energy, a simple expression obtains,
\begin{equation}
\Delta\pot=\frac{1}{2}\loge\Bigl(
\frac{\rho_-(\infty)}{\rho_+(\infty)}\cdot
\frac{\rho_+(-\infty)}{\rho_-(-\infty)}\Bigr).\label{eq:donnan}
\end{equation}
This method of calculating the junction potential was used in
Ref.~\cite{Warren1}.  

One question remains: what should be used for the chemical potentials
in these calculations?  The simplest answer is to compute the chemical
potentials from Eq.~\eqref{eq:mu}, setting $\pot=0$ and using the bulk
densities corresponding to either one of the coexisting phases.  This
works because global charge neutrality means Eq.~\eqref{eq:gamma} for
the surface tension is unaffected by the value of $\pot$ in
Eq.~\eqref{eq:mu}.  Hence we are free to set $\pot=0$ in either of the
coexisting phases.

\begin{figure}
\begin{center}
\includegraphics{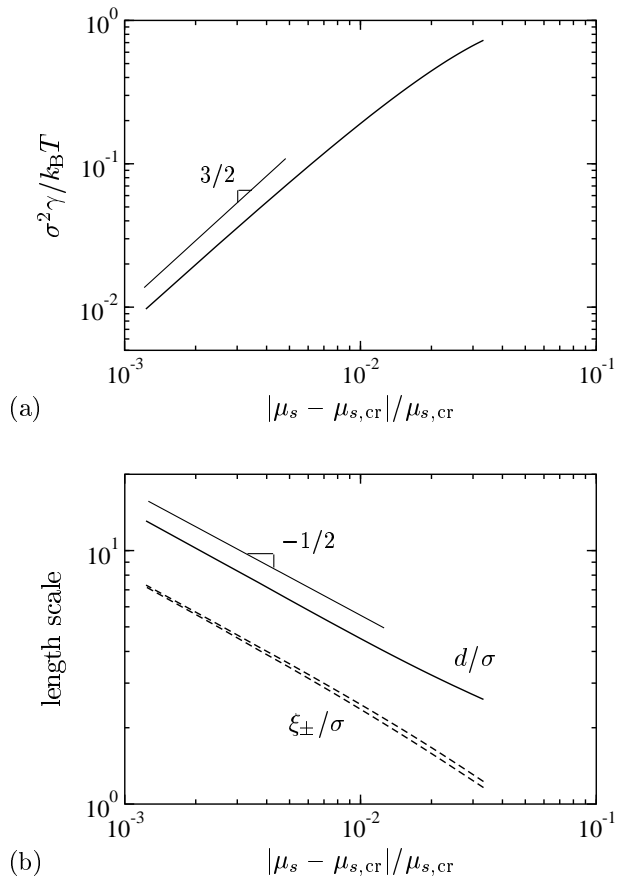}
\end{center}
\caption[?]{(a) Surface tension as a function of salt chemical
potential.  (b) Interface width (solid line) and correlation lengths
(dashed lines) as a function of salt chemical potential.  In both, the
salt chemical potential is expressed as a normalised distance from the
upper critical point.\label{fig:scaling}}
\end{figure}

I now turn to the results.  Fig.~\ref{fig:ionprof} shows
representative density profiles for the macroion and small ions
through the interface between the coexisting phases, corresponding to
the highlighted tie line in Fig.~\ref{fig:bino}.  The profiles
interpolate smoothly between the coexisting bulk densities.
Fig.~\ref{fig:fieldprof} shows the detailed electrical structure at
the interface.  The upper plot shows that the charge density
$\rhoz=Z\rho_M+\rho_+-\rho_-$ has a dipolar structure.
Correspondingly there is a localised electric field, shown in the
middle plot, and a smooth jump of $\Delta\pot\approx20.5\,\mathrm{mV}$
in the electrostatic potential, shown in the lower plot.  This is the
junction potential which can also be calculated directly from the
coexisting bulk densities as in Eq.~\eqref{eq:donnan}.  This
electrical structure is in accord with general expecatations for
charged systems \cite{Sluckin, NabuNGED}.

Fig.~\ref{fig:omprof} shows the grand potential density and the
electrostatic component thereof---the second term of
Eq.~\eqref{eq:omdens}---as a function of distance through the
interface.  For this particular case the area gives $\gamma \approx
0.727\times(\kT/\sigma^2)$.  The order of magnitude of this should not
come as a surprise since $\sigma$ and $\kT$ are the only relevant
length and energy scales in the problem.  Inserting actual values,
$\gamma\sim 0.3\,\mu\mathrm{N}\,\mathrm{m}^{-1}$, which is typical for
for soft matter interfaces \cite{surftens}

Fig.~\ref{fig:scaling} shows how the surface tension and interface
width vary as one approaches the upper critical point in
Fig.~\ref{fig:bino}.  The width $d$ is defined operationally as
$d^2=\langle x^2\rangle-\langle x\rangle^2$, where $\langle
\dots\rangle=\int_{-\infty}^{\infty}(\dots)\,p(x)\,dx/
\int_{-\infty}^{\infty}p(x)\,dx$, with
$p(x)=|\omega(x)-\omega(\pm\infty)|^2$.  These results are obtained by
repeating the calculations underlying
Figs.~\ref{fig:ionprof}--\ref{fig:omprof} for a sequence of tie lines
approaching the critical point.  They are reported as a function of
the distance from the critical point, expressed in terms of a
normalised salt chemical potential.  Fig.~\ref{fig:scaling}(b) also
shows the correlation lengths $\xi_{\pm}$ in the coexisting phases
determined from the exponential decay of the density profiles into the
bulk phases (see Appendix B).  As the critical point is
approached, these approach each other, and diverge in the same way as
the interface width.  Fig.~\ref{fig:scaling} reveals that the surface
tension and length scales are in accord with expected scaling
behaviour for a mean-field theory \cite{WidRowl}.

What happens at the lower critical point in Fig.~\ref{fig:bino}
though?  The next section shows that this is a non-trivial question
with perhaps an unexpected answer.  In the calculations in the current
section, I have assumed that the interface profiles smoothly
interpolate between the coexisting phases.  Indeed, this is the basis
of the numerical method detailed in Appendix B.  However,
such an approach rules out the possibility of oscillatory behaviour in
the density profiles (or to be precise, the numerical methodology is
inappropriate for this scenario).  At lower salt concentrations
though, one can enter a region where oscillatory behaviour is
expected.  These considerations are made mathematically precise in the
next section.

\begin{figure}
\begin{center}
\includegraphics{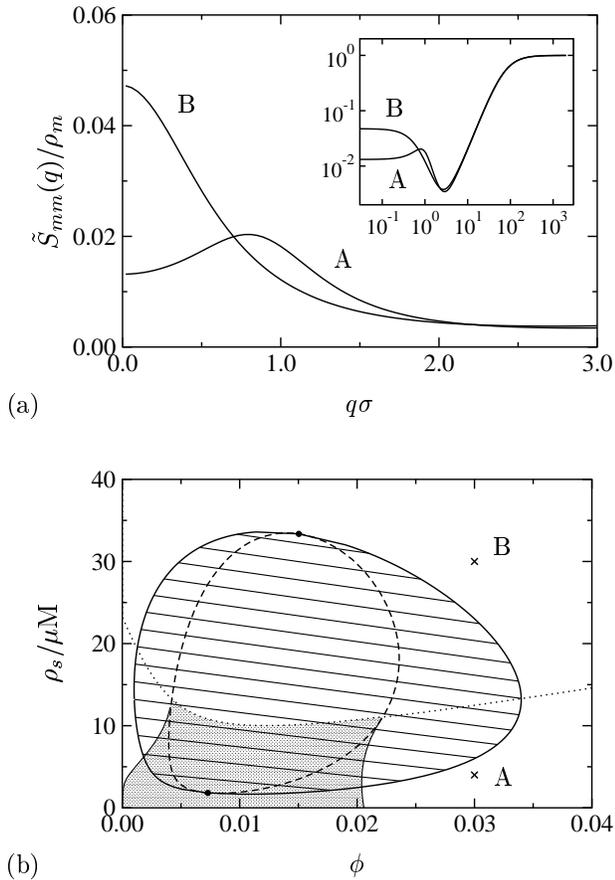}
\end{center}
\caption[?]{(a) Macroion structure factors at $\vf=0.03$, and
$\rho_s=4\,\muM$ (A) and $30 \,\muM$ (B).  The inset shows the same
curves in a double-logarithmic plot.  The normalisation is such that
$\tilde S_{mm}/\rho_m\to1$ as $q\to\infty$.  (b) Phase diagram
augmented by the spinodal line (dashed), the \Lif\ line (dotted), and
the region where the macroion structure factor diverges at a non-zero
wavevector (shaded).\label{fig:sfactor}}
\end{figure}

\section{Structure factors}
\label{sec:sf}
The structure factors in a homogeneous system can be determined from a
density functional theory (DFT) by functional differentiation
\cite{EvansDFT}.  Where accurate structure factors are already
known, typically from a combination of simulation and integral
equation approaches, this can be used to constrain the DFT.  In the
present case for example, one could try to constrain $w(r)$ in
Eq.~\eqref{eq:wr}.  However accurate structure factors are not known
for this problem, and furthermore the DFT has been constructed to
include only the macroion self energy.  Thus it does not make sense to
constrain the DFT and the present section simply reports the structure
factors that are predicted from the theory as given in
Eqs.~\eqref{eq:dft}--\eqref{eq:wr}.

The structure factor matrix is \cite{HansMac, MarchTosi}
\begin{equation}
\tilde S_{ij}(q)=\rho_i\delta_{ij}+
\rho_i\rho_j\tilde h_{ij}(q)
\label{eq:sij}
\end{equation}
where $i$ and $j$ run over $\{m,+,-\}$ and $\tilde h_{ij}(q) = \int
\dthree\rvec\, e^{-i\qvec\cdot\rvec}\, h_{ij}(r)$ is the Fourier
transform of the pair correlation functions $h_{ij}(r) = g_{ij}(r) -
1$.  Reciprocal space quantities will be denoted by a tilde.  The bulk
densities $\rho_i$ are constants, fixed by the choice of state point.
Deviations away from these will be denoted by $\Delta\rho_i$.
Eq. \eqref{eq:sij} uses the normalisation $\tilde
S_{ij}(q)\to\rho_i\delta_{ij}$ as $q\to\infty$, which simplifies some
of the expressions below \cite{MarchTosi}.

To obtain the structure factor matrix, start by defining the
real-space function
\begin{equation}
S^{-1}_{ij}(|\rvec-\rvec'|)
=\frac{1}{\kT}\Bigl(\frac{\delta^2\!F}
{\delta\rho_i(\rvec)\delta\rho_j(\rvec')}\Big)_{\rho_i(\rvec)\to\rho_i}
\label{eq:sm1}
\end{equation}
where $F$ is the full free energy.  The limit of a homogeneous system
is taken after the functional differentiation step so that
$S^{-1}_{ij}$ only depends on $|\rvec-\rvec'|$ as indicated.
Transforming to reciprocal space, one can show that
\begin{equation}
\tilde S^{-1}_{ij}(q) = {\textstyle\int}\dthree\rvec\,e^{-i\qvec\cdot\rvec}
\,S^{-1}_{ij}(r)
\end{equation}
is simply the matrix inverse of $\tilde S_{ij}$,
\begin{equation}
\sum_j\tilde S_{ij}\tilde S^{-1}_{jk} = \delta_{ik}.\label{eq:matinv}
\end{equation}
These results follow by combining the Ornstein-Zernike relation for a
multicomponent mixture in reciprocal space, $\tilde h_{ij}=\tilde
c_{ij}+\sum_k\rho_k\tilde c_{ik}\tilde h_{jk}$ where $c_{ij}$ are the
direct correlation functions \cite{HansMac}, with the DFT result that
$c_{ij}=-(1/\kT)\delta^2\!\Fex / \delta\rho_i\delta\rho_j$ where
$\Fex$ is the excess free energy \cite{EvansDFT}.

The route to the structure factors offered by
Eqs.~\eqref{eq:sm1}--\eqref{eq:matinv} is based on `classical'
arguments \cite{HansMac}.  One can also make the connection via field
theoretical methods.  Expanding the free energy functional to second
order gives
\begin{equation}
\frac{\Delta F}{\kT}=\frac{1}{2}\int\!
\dthree\rvec\,\dthree\rvec'\sum_{ij}
\Delta\rho_i(\rvec)\Delta\rho_j(\rvec')
S^{-1}_{ij}(|\rvec-\rvec'|),
\end{equation}
where $S^{-1}_{ij}$ is defined by Eq.~\eqref{eq:sm1}.
It follows that \cite{DoiEdwards}
\begin{equation}
\langle\Delta\rho_i(\rvec)\Delta\rho_j(\rvec')\rangle
= S_{ij}(|\rvec-\rvec'|)\label{eq:sij2}
\end{equation}
where $S_{ij}(r)=\int \dthree\qvec/(2\pi)^3 e^{i\qvec\cdot\rvec}\,
\tilde S_{ij}(q)$ is the structure factor matrix expressed as a real
space quantity.  Although care has to be taken at the point
$\rvec=\rvec'$, one can easily show that the density-density
correlation function on the left hand side of Eq.~\eqref{eq:sij2} is
the same as the Fourier transform of the right hand side of
Eq.~\eqref{eq:sij}.

The \SL\ moment conditions constrain the behaviour of the structure
factors in reciprocal space in a particularly clear manner
\cite{StillLov, EvSl, Martin, StaBad, LeeFish2}.  Firstly, the
zeroth-moment conditions express perfect screening and are
$\int\dthree\rvec\, \sum_i z_i \rho_i g_{ij}(r) = - z_j$ for
$j=\{m,+,-\}$.  Using charge neutrality and assuming the structure
factors are regular at $q=0$, one can easily show that this implies
\begin{equation}
\textstyle\sum_iz_i\tilde S_{ij}(\qvec)=O(q^2).\label{eq:sl0}
\end{equation}
The
second-moment condition is $\int\dthree\rvec\,r^2\,\sum_{ij}
z_iz_j\rho_i\rho_jg_{ij}(r) =-{3}/{(2\pi\lB)}$.  This
constrains the long wavelength behaviour of the charge-charge
structure factor,
\begin{equation}
{\textstyle\sum_{ij}z_iz_j\tilde S_{ij}}(\qvec)
=\frac{q^2}{(4\pi\lB)}+O(q^4).\label{eq:sl2}
\end{equation} 
In real space, this means that $\langle\Delta\rhoz(\rvec)
\Delta\rhoz(\rvec')\rangle \sim \lB/|\rvec-\rvec'|$ for
$|\rvec-\rvec'|\to\infty$.  Thus charge density fluctuations vanish
with the Coulomb law at large distances, corresponding to the fact
that the electrostatic energy dominates in the free energy for
long-wavelength density fluctuations unless they happen to be
charge-neutral \cite{note3a}.

I now apply the formalism of Eqs.~\eqref{eq:sm1}--\eqref{eq:matinv} to
the present DFT defined in Eqs.~\eqref{eq:dft}--\eqref{eq:wr}.  The
result for the inverse structure factor matrix in reciprocal space can
be written as
\begin{equation}
\tilde S^{-1}_{ij}=\tilde T^{-1}_{ij}
+\frac{4\pi\lB z_iz_j}{q^2}\label{eq:sijfull}
\end{equation}
where first term comes from the ideal and correlation contributions to
the free energy and the second term from the mean field
electrostatics.  The first term is in detail
\begin{equation}
\begin{array}{l}
\tilde T^{-1}_{ij}={\delta_{ij}}/{\rho_i}
+\rho_mZ^2\pi^2\lB^3\sigma^3\,h_1(\sigma\kappa,\sigma q)\Delta'_{ij}\\[9pt]
{}\hspace{10em}-Z^2\pi\lB^2\sigma\,h_2(\sigma\kappa,\sigma
q)\Delta''_{ij}.
\end{array}\label{eq:tij}
\end{equation}
where the functions $h_{1,2}(x=\sigma\kappa, y=\sigma q)$ are
\begin{equation}
h_1=\frac{8e^{-\alpha y^2/2}(2+3x)}{(x^3(x+2)^3)},\quad
h_2=\frac{4e^{-\alpha y^2/4}}{(x(x+2)^2)},
\label{eq:hfns}
\end{equation}
and the matrices are 
\begin{equation}
\begin{array}{ll}
\Delta'_{mm}=\Delta'_{m\pm}=0,&{}\quad \Delta'_{\pm\pm}=1,\\[9pt]
\Delta''_{mm}=\Delta''_{\pm\pm}=0,&{}\quad \Delta''_{m\pm}=1.
\end{array}\label{eq:dij}
\end{equation}
The $y$-dependence ($y=\sigma q$) in Eq.~\eqref{eq:hfns} arises from
the Fourier transform of the weight function of Eq.~\eqref{eq:wr}.
Note that the point model alluded to in section \ref{sec:model}
corresponds to the limit $\alpha\to0$ in Eqs.~\eqref{eq:hfns}.  In this
limit, the theory becomes ill-defined since $\tilde S_{ij}(q)$ does
not have the correct limiting behaviour as $q\to\infty$.  This was the
original technical reason for introducing the smoothing kernel.

For any given state point and value of $q$,
Eqs.~\eqref{eq:sijfull}--\eqref{eq:dij} define $\tilde S^{-1}_{ij}$
which can be inverted numerically to find all components of the
structure factor matrix.  A partial solution can be obtained
analytically in terms of the subsidiary matrix $\tilde T_{ij}$,
\begin{equation}
\tilde S_{ij}=\tilde T_{ij}
-\frac{4\pi\lB \sum_{kl}z_kz_l\tilde T_{ik}\tilde T_{jl}}
{q^2+4\pi\lB\sum_{kl}z_kz_l\tilde T_{kl}}
\end{equation}
From this one can readily prove that $\tilde S_{ij}$ exactly satisfies
the \SL\ moment conditions in Eqs.~\eqref{eq:sl0} and \eqref{eq:sl2}
above.

Another result follows from the dominance of the ideal contribution
over the correlation contribution at low densities.  In the limit
$\rho_i\to0$ one finds $\tilde T_{ij}\to\rho_i\delta_{ij}$ and
\begin{equation}
\tilde S_{ij}\to\rho_i\delta_{ij}
-\frac{4\pi\lB z_iz_j\rho_i\rho_j}{q^2+4\pi\lB\sum_kz_k^2\rho_k}.
\label{eq:lowd}
\end{equation}
This is in fact exactly in accordance with the \DebH\ limiting law at
low densities.  To see this, note that
$\lambda=(4\pi\lB\sum_kz_k^2\rho_k)^{-1/2}$ is the Debye screening
length defined to include \emph{all} ionic species.  Thus in real
space, Eqs.~\eqref{eq:sij} and \eqref{eq:lowd} indicate that
$h_{ij}=-z_iz_j(\lB/r)e^{-r/\lambda}$, in correspondence with the
\DebH\ limiting law.

It is clear that the moment conditions and the \DebH\ limiting law
behaviour follow from the construction of the DFT to include a
mean-field contribution separately from the correlation term.  This
construction is in turn motivated by the expected behaviour of the
direct correlation functions $c_{ij}(r)$ at $r\to\infty$, as Evans and
Sluckin have described \cite{EvSl}.  The form of the correlation term
is unimportant, so long as it is regular both at $q\to0$ and
$\rho_i\to0$.

For the remaining part, I now focus on the macroion structure factor
$\tilde S_{mm}$.  Note that the theory includes the macroion-macroion
electrostatic interaction explicitly in the mean field term, and an
additional indirect interaction in the correlation term.  The
computation of $\tilde S_{mm}$ reveals the combined effect of these
macroion \emph{interactions} on the macroion \emph{correlations}.

Typically $\tilde S_{mm}$ has a `hole' in reciprocal space for
$q\sigma\alt1$.  This corresponds to the macroion electrostatic
repulsions.  Within the correlation hole though, there is additional
structure.  This becomes particularly important in the vicinity of the
phase separation region.  Two kinds of behaviour are possible: at
higher salt concentrations $\tilde S_{mm}$ rises to a maximum as
$q\to0$, or at lower salt concentrations $\tilde S_{mm}$ acquires a
peak at some $q^*>0$.  In the phase diagram, the two alternatives are
separated by a (macroion) `\Lif\ line' \cite{ALev}, defined to be the
locus of points for which $\partial\tilde
S_{mm}/\partial(q^2)|_{q=0}=0$.  Fig.~\ref{fig:sfactor}(a) shows the
two behaviours for a pair of typical state points above and below the
\Lif\ line, and Fig.~\ref{fig:sfactor}(b) shows the \Lif\ line
superimposed on the bulk phase behaviour.

Also shown in Fig.~\ref{fig:sfactor}(b) is the spinodal line computed
from the bulk free energy in Eq.~\eqref{eq:bfe} of section
\ref{sec:bulk}.  One can check that $\tilde S_{mm}(q=0)$ diverges on
this spinodal line; in fact all the $q=0$ components of the structure
factor matrix diverge because the determinant of $\tilde S^{-1}_{ij}$
vanishes.  For salt concentrations above the \Lif\ line, this
divergence at $q=0$ can be accommodated within the general behaviour
of the structure factor.  Of course, state points within the binodal
are metastable so the divergence is strictly only visible as the upper
critical point is approached.  The fact that the structure factors
diverge on the spinodal line is no coincidence, since thermodynamic
consistency by the compressibility route is assured for a DFT
\cite{note3}.

What happens at salt concentrations below the \Lif\ line?  Here, the
peak in $S_{mm}$ at $q^*>0$ is found to diverge \emph{before} the bulk
spinodal line is reached.  The shaded area in
Fig.~\ref{fig:sfactor}(b) shows the region where this occurs.  A
divergence at a non-zero wavevector is indicative of microphase
separation \cite{microphase}.  In this case one would expect a
charge-density-wave (CDW) phase to appear \cite{NabuNP1, NabuNP2}.
The shaded region extends below the binodal for bulk phase separation,
so the CDW phase should be observable in this part of the phase
diagram.  In fact the CDW phase will be found whenever the lower
critical point lies below the \Lif\ line.  The general idea that a
critical point in a charged system can be replaced by a CDW phase was
advanced by Nabutovskii, Nemov and Peisakhovich \cite{NabuNP1,
note4b}.

The location of the \Lif\ line depends on the parameter $\alpha$ which
sets the range of the smoothing kernel $w(r)$ in Eq.~\eqref{eq:wr}.
If $\alpha\alt0.40$ the \Lif\ line moves upwards past the upper
critical point, which would then be expected to be replaced by a CDW
phase too.  On the other hand if $\alpha\agt3.6$, the \Lif\ line moves
downwards past the lower critical point.  These critical values of
$\alpha$ only depend on the coefficient of $q^2$ in the expansion of
the Fourier transform of $w(r)$ about $q=0$.  

The \Lif\ line discussed here pertains to the macroion structure
factor.  Although slightly different \Lif\ lines are expected for each
component of the structure factor matrix, the locus of state points
where the peak diverges (either on the spinodal or on the boundary of
the CDW phase) should be the same for all components.

Whilst the \Lif\ line line marks an obvious change in the behaviour of
$\tilde S_{mm}$, the cross-over from monotonic to damped oscillatory
asymptotic decay of the correlation functions $h_{ij}(r)$ is
determined by Kirkwood or Fisher-Widom lines in the phase diagram
\cite{KFWline, EvLCHH, LCEv}.  The difference between these
is rather subtle \cite{LCEv, note4}, and one might loosely cover both
possibilities by the phrase `\KFW' (KFW) line.  The importance of the
KFW line lies in the fact that it also governs the asymptotic decay of
the interface density profiles, which behave in the same way as
$h_{ij}$ \cite{EvLCHH}.  Thus the calculations reported in section
\ref{sec:profile} above, which assume that there is no oscillatory
behaviour in the density profiles, requires as a necessary minimum
that the coexisting bulk densities both lie above the KFW line.  The
location of the KFW line is governed by the poles of $\tilde
S_{ij}(q)$ in the complex $q$ plane, which are either purely imaginary
or occur as complex conjugate pairs, and are the same for all
components of $\tilde S_{ij}$ \cite{EvLCHH}.  If the pole nearest the
real $q$-axis is purely imaginary, then monotonic decay is expected;
conversely if a pair of complex conjugate poles is nearest the real
$q$-axis, then damped oscillatory decay is expected \cite{LCEv}.
Determination of the KFW line is a hard numerical problem and has not
been attempted for the present DFT.  However the presence of a peak in
$\tilde S_{mm}(q)$ on the real $q$-axis at $q=0$, or at $q^*>0$, ought
to be indicative of whether the pole nearest the real $q$-axis is, or
is not, purely imaginary.  Thus the \Lif\ line should serve as a guide
to the location of the KFW line.  In section \ref{sec:profile}
therefore, care was taken to make sure that the coexisting bulk
densities lie well above the \Lif\ line.

\section{Discussion}
The paper presents a density functional theory (DFT) for a macroion
suspension.  The excess free energy corresponds to the macroion self
energy evaluated using \DebH\ theory.  These approximations render
theory tractable without losing the basic phenomenology which
resembles that of other studies.  The advantage of a DFT is that one
can compute the interface structure and surface tension between
coexisting phases.  The results are in accord with expectations from
previous work \cite{Warren1}.  In particular, the electrical structure
of the interface gives rise to a junction potential analogous to the
Donnan potential across a semi-permeable membrane.  This arises from
an electric dipole moment density (per unit area of interface), which
appears because charge neutrality is locally violated in the vicinity
of the interface.  The surface tension is found to be of the order
$\kT/\sigma^2$.

Structure factors can be computed from the DFT.  These are found to
obey the Stillinger-Lovett moment conditions, although this is not a
stringent test of the theory.  The structure factors reveal an
interesting phenomenon, namely that oscillatory behaviour can appear
in the (direct) correlation functions, particularly at low ionic
strength.  Indeed there may be regions of microphase separation in the
vicinity of the critical points, corresponding to the appearance of a
charge-density-wave (CDW) phases.  This phenomenon is peculiar to
asymmetric charged systems \cite{NabuNP1}, and is strictly absent in
symmetric systems such as the restricted primitive model.  In this
respect, the possibility of CDW phases is correlated with the
appearance of the junction potential, which is also strictly absent in
symmetric systems \cite{Warren3}.  Given the approximate nature of the
DFT, only certain aspects of the present analysis might be expected to
survive in a full treatment.  One of these is an upturn in macroion
structure factor at small $q$, even in the absence of a true
miscibility gap.  This would reflect an increased osmotic
compressibility in this region of the phase diagram.  Another
expectation is the possible appearance of the CDW phases, although it
might be difficult to disentangle these from the ordered (crystal)
phases that are expected for a macroion suspension at sufficiently
strong electrostatic coupling.

The macroion self energy depends on the local ionic strength, but on
both physical and technical grounds it is found necessary to introduce
the notion of smoothing or smearing---the dependency should be on the
ionic strength averaged over the vicinity of the macroion.  Here a
completely phenomenological approach has been taken to construct the
details of the DFT.  Other choices could be made, or indeed more rigor
could be introduced, such as additional requirements for internal
consistency \cite{note5}.  Tests indicate though that the general
phenomenology (electrical structure of interface, gross behaviour of
structure factors) is found to be insensitive to the details of the
model at this point.

\begin{acknowledgments}
I thank R. Evans and A. S. Ferrante for useful discussions.
\end{acknowledgments}

\appendix

\section{Correction to Ref.~\cite{Warren1}}
\label{sec:chan}
Chan \cite{Chan1} has remarked that an excluded volume contribution
was omitted in the theory of Ref.~\cite{Warren1}.  This appendix
describes the missing term.  The error occurs in going from Eq.~(3) to
Eq.~(7) of Ref.~\cite{Warren1} where the omitted contribution arises
from the fact that $h_{m\pm}(r)=g_{m\pm}(r)-1=-1$ for $r<\sigma/2$.
In terms of the microion-macroion interaction energy,
$E_{\mathrm{ms}}/(V\kT)$, the omitted contribution is
\begin{equation}
\begin{array}{l}
\displaystyle
\rho_m\int_{|\rvec|<\sigma/2}\!\!\dthree\rvec\,
\frac{Z\lB}{r}\,[\rho_+h_{m+}(r)-\rho_-h_{m-}(r)]\\[15pt]
\displaystyle
{}\quad=-\rho_m(\rho_+-\rho_-)\int_0^{\sigma/2}\!\!
4\pi r^2dr\,\frac{Z\lB}{r}\\[15pt]
\displaystyle
{}\quad=+\frac{\pi Z^2\lB\rho_m^2\sigma^2}{2}\quad
\hbox{(using $\rho_+-\rho_-=-Z\rho_m$).}
\end{array}
\end{equation}

This contribution is a positive, increasing function of $\rho_m$, and
has the tendency to stabilise the system against phase separation
(because it is an athermal excluded volume term, it passes unscathed
through the thermodynamic integration step needed to calculate the
contribution to the free energy).  If the calculations of
Ref.~\cite{Warren1} are repeated with this contribution included, it
is found that the basic phenomenology is still the same, except that
the miscibility gap in the $(\rho_m, \rho_s)$ plane does not appear
until somewhat larger values of $Z\lB/\sigma$.  Fig.~8
shows the new results in comparison with those reported in Table II of
Ref.~\cite{Warren1}.  The new calculation indicates that phase
separation is observed in an even narrower window of parameter space
for which the \DebH\ linearisation approximation might be admissible
than was found in the earlier work.  This can be taken to indicate
that the self-energy mechanism may not be sufficiently powerful to
drive phase separation by itself, as discussed in the introduction.

\begin{figure}
\begin{center}
\includegraphics{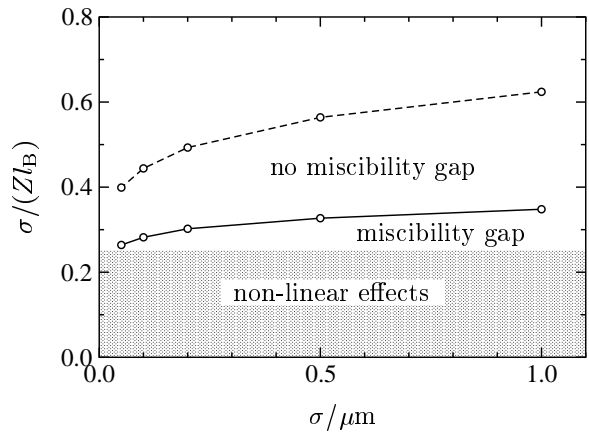}
\end{center}
\caption[?]{State diagram showing where a miscibility gap is found for
the full theory of Ref.~\cite{Warren1} including the omitted term
(solid line), compared to the original results (dashed line).  The
shaded region shows where $Z\agt 4\sigma/\lB$, which is one possible
criterion for the acceptability of the \DebH\ approximation for the
polarisation energy \cite{Warren1}.\label{fig:chan}}
\end{figure}

\section{Numerical approach}
\label{sec:num}
The task is to find density profiles $\rho_i(x)$ which minimise the
grand potential in Eq.~\eqref{eq:omega}.  The most accurate method is
to solve the integral equations for the profiles in
Eq.~\eqref{eq:horrid}.  However, this is hard.  An alternative is to
adopt a variational approach in which $\Omega$, or $\gamma$ in
practice, is minimised with respect to parameters in trial functions
which specify the density profiles \cite{SmithOrosz}.  This is the
approach that has been taken here.

The ion density profiles have to satisfy a sum rule since the
potential difference $\Delta\pot=\pot(\infty)-\pot(-\infty)$ is fixed
by the coexisting bulk densities as described in section
\ref{sec:profile}.  One can replace one of the ion density profiles by
$\pot(x)$ to ensure this sum rule is automatically satisfied.  In the
present case, a choice was made to use the set
$\{\rho_m,\rho_+,\pot\}$ as a basis with $\rho_-$ derived analytically
from the Poisson equation, $\rho_- = Z\rho_m + \rho_- -
(d^2\pot/dx^2)/(4\pi\lB)$.  The first integral of the Poisson equation
shows that one can additionally ensure global charge neutrality by
making sure that $d\pot/dx\to0$ as $|x|\to\infty$.  Once the $\rho_i$
are known, the average ionic strength $\overline\rho_I$ and the
surface tension $\gamma$ are determined numerically by quadratures.

To represent the basis set $\{\rho_m,\rho_+,\pot\}$, three copies of
the function
\begin{equation}
\begin{array}{l}
\displaystyle f(x;\xi_\pm,\{a\})=\frac{a_- e^{x/\xi_+}-a_+ e^{-x/\xi_-}}
{a_-a_+ + a_- e^{x/\xi_+}+a_+ e^{-x/\xi_-}}\\[12pt]
{}\hspace{10em}+\sum_{r=1}^N a_r H_r(x/\xi)
\end{array}
\end{equation}
are introduced.  In this, the $H_r$ are Hermite functions, with
$\xi=2/(1/\xi_-+1/\xi_+)$ used to scale the argument.  Each copy of
$f$ is parametrised by the correlation lengths $\xi_\pm$ and amplitude
set $\{a\}$, and has the properties that $f\to\pm(1-a_\pm e^{\mp
x/\xi_\pm})$ as $x\to\pm\infty$.  One copy of $f$ is assigned to each
member of $\{\rho_m,\rho_+,\pot\}$, and is scaled and shifted to match
the limiting values at $|x|\to\infty$, for example
$\rho_m=\rho_m(-\infty)(1-f)/2+\rho_m(\infty)(1+f)/2$ (for the
electrostatic potential, one can set $\pot(-\infty)=0$ and
$\pot(\infty)=\Delta\pot$).  The three copies of $f$ have different
amplitude sets $\{a\}$ but share common values for $\xi_\pm$ since the
asymptotic decay of the density profiles into the bulk phases is
expected to be governed by a bulk correlation length---it is these
values of $\xi_\pm$ that are reported in Fig.~\ref{fig:scaling}(b).  A
finite set of $N$ Hermite functions has been included in each copy of
$f$ to allow for an arbitrary structure at the interface.  In practice
the minimisation problem is well behaved only if the density profiles
smoothly interpolate between the bulk values, for which case typically
$N=3$--6 Hermite functions are needed to achieve convergence in
$\gamma$ to an accuracy of the order 1\%.  At this point, the
interface problem has been reduced to a multivariate minimisation over
the three copies of the amplitude set $\{a\}$ plus the correlation
lengths $\xi_\pm$.  Numerical minimisation of $\gamma$ with respect to
these parameters is then undertaken by standard methods \cite{NumRec}.



\bibliography{ipaper}

\begin{thebibliography}{76}
\bibitem[{not({\natexlab{a}})}]{note1}
\bibinfo{note}{This expression includes the bare self energy of the macroion,
  although this is unimportant unless the macroion charge is allowed to vary
  \cite{DBLev}. More complicated expressions for the self energy are available
  \cite{Warren1}, but it has been checked that the use of these does not alter
  the basic phenomenology.}
\bibitem[{not({\natexlab{b}})}]{note1b}
\bibinfo{note}{It has been checked that the phenomenology is robust with
  respect to the choice of $w(r)$.}
\bibitem[{not({\natexlab{d}})}]{note2}
\bibinfo{note}{Eq.~\eqref{eq:pd2} limits to $\pot\rhoz/2$ which is different
  from Eq.~\eqref{eq:bfe}. This is because the constraint term in
  Eq.~\eqref{eq:bfe} is inserted by hand and does not correspond to a charging
  process. The difficulty is more apparent than real though, since it
  disappears when the chemical potentials are computed. Also note that $\pot$
  is not the true mean electrostatic potential because it ignores the
  electrical structure of the double layers around the macroions.}
\bibitem[{not({\natexlab{f}})}]{note3}
\bibinfo{note}{One would not expect consistency by the energy or virial routes
  because these invoke the pair potentials. Whilst the pair potentials are of
  course specified in the primitive model, the DFT itself is approximate.}
\bibitem[{not({\natexlab{h}})}]{note4}
\bibinfo{note}{Roughly, the period of the oscillations diverges as one
  approaches a Kirkwood line, but not as one approaches a Fisher-Widom line;
  see Ref.~\cite{LCEv}.}
\bibitem[{not({\natexlab{i}})}]{note5}
\bibinfo{note}{For example, it has not been checked that $g_{ij}(r)\ge0$. Also,
  one could require that the $g_{ij}(r)$ computed from the structure factors
  are the same as those that can be computed by solving the inhomogeneous
  problem of the density profiles in an appropriately chosen external potential
  \cite{EvansDFT}. Such requirements would introduce additional numerical
  complexity into the problem though.}
\end{thebibliography}

\end{document}